\documentclass[aps,prc,twocolumn,superscriptaddress,amsfonts,amsmath,amssymb,showpacs,floatfix]{revtex4-1}
\usepackage[pdftex]{graphicx} 
\usepackage{hyperref,longtable,dcolumn}
\usepackage{color}
\usepackage[dvipsnames,svgnames,x11names,hyperref]{xcolor}
\hypersetup{colorlinks,citecolor=[RGB]{132 112 255},filecolor=[RGB]{220 20 60},linkcolor=[RGB]{191 62 255},urlcolor=[RGB]{138 43 226}}
\usepackage{float}
\usepackage{footnote}

\begin{document}


\title{Neutron-hole strength in \emph{N} = 81 nuclei}

\author{A. M. Howard}
\altaffiliation[Current address: ]{ FRM-II Heinz Maier-Leibnitz Research Neutron Source,
Technical University of Munich, Lichtenbergstrasse 1
85748 Garching
Germany.}
\affiliation{Department of Physics and Astronomy, University of Manchester, Manchester M13 9PL, United Kingdom}

\author{S. J. Freeman}
\email[Correspondence to: ]{sean.freeman@manchester.ac.uk}
\affiliation{Department of Physics and Astronomy, University of Manchester, Manchester M13 9PL, United Kingdom}
\author{D. K. Sharp}
\affiliation{Department of Physics and Astronomy, University of Manchester, Manchester M13 9PL, United Kingdom}

\author{T. Bloxham}
\affiliation{Physics Department, University of California, Berkeley, California 94720, USA}
\affiliation{Lawrence Berkeley National Laboratory, Berkeley, California 94720, USA}

\author{J. A.~Clark}
\affiliation{Physics Division, Argonne National Laboratory, Argonne, Illinois 60439, USA}

\author{C.~M.~Deibel}
\altaffiliation[Current address: ]{Department of Physics and Astronomy, Louisiana State University, Baton Rouge, LA 70803,
USA.}
\affiliation{Physics Division, Argonne National Laboratory, Argonne, Illinois 60439, USA}

\affiliation{Joint Institute for Nuclear Astrophysics, Michigan State University, East Lansing, Michigan 48824, USA}

\author{B. P. Kay}
\affiliation{Physics Division, Argonne National Laboratory, Argonne, Illinois 60439, USA}

\author{P. D. Parker}
\affiliation{A. W. Wright Nuclear Structure Laboratory, Yale University, New Haven, Connecticut 06520, USA}

\author{J. P. Schiffer}
\affiliation{Physics Division, Argonne National Laboratory, Argonne, Illinois 60439, USA}

\author{J. S. Thomas}
\affiliation{Department of Physics and Astronomy, University of Manchester, Manchester M13 9PL, United Kingdom}

\date{\today}

\begin{abstract}
A systematic study of neutron-hole strength in the $N=81$ nuclei $^{137}$Ba, $^{139}$Ce, $^{141}$Nd and $^{143}$Sm is reported. The single-neutron removal reactions ($p$,$d$) and ($^3$He,$\alpha$) were measured at energies of 23 and 34~MeV, respectively. Spectroscopic factors were extracted from measured cross sections through a distorted-wave Born approximation analysis and centroids of single-particle strength have been established. The change in these centroid energies as a function of proton number have been compared to calculations of the monopole shift for the $s_{1/2}$ and $h_{11/2}$ orbitals, where the majority of the strength has been observed. Significant fragmentation of strength was observed for the $d$ and $g_{7/2}$ orbitals, particularly for the latter orbital which is deeply bound, with summed strengths that indicate a significant amount lies outside of the measured excitation energy range.

\end{abstract}


\maketitle


\section{Introduction}

The description of atomic nuclei in terms of constituent nucleons moving within a mean-field potential is the basis of the shell model, and consequently, much of our understanding of nuclear structure. Over the past decade or so, evidence has emerged indicating that, when moving away from stability into exotic systems, the ordering of single-particle levels evolves as a function of  proton and neutron number to the extent that the gaps between levels that correspond to shell and sub-shell closures are found to alter. Significant attention has been paid to these phenomena in the literature, which has motivated a careful reexamination of how the interaction between valence protons and neutrons drives such evolution. On moving through a series of isotopes or isotones, the changing single-particle occupancies of one type of nucleon alters the overall effect of interactions with a nucleon of the other type, thus changing its effective single-particle energy.  It appears that in some cases both the central and tensor components of the nucleon-nucleon interaction need to be considered carefully in order to reproduce the observed changes in single-particle structure \cite{Otsuka1, Otsuka2, Smirnova}.

It is therefore interesting to carefully reexamine the trends in single-particle states near the line of $\beta$ stability, particularly where changes can be tracked across a range of proton-neutron ratios. Such experimental measurements are often easier and tend to yield more detailed information compared to studies with radioactive beams, which are performed with inevitably lower beam intensities. In many experiments with stable beams, centroids of single-particle strength can  be constructed from the observation of several different excited states populated by transfer of a nucleon to the same orbital and used to estimate its effective single-particle energy. 

\begin{figure}[htb]
\centering
\includegraphics[scale=0.80]{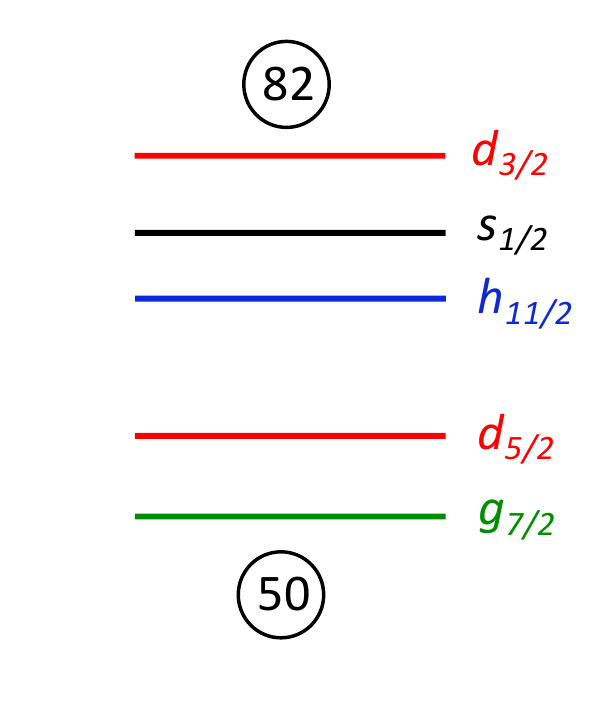}
\caption{\label{Sch} Schematic level diagram of the single-particle orbitals near stability for the shell between $N=50$ and $N=82$.}
\end{figure}

Several studies have been performed recently using consistent approaches to both experimental and analytical methods  that have highlighted the detailed trends in single-particle orbitals in near stable nuclei. These include studies of high-$j$ proton states outside of stable Sn cores \cite{Sn-at}; untangling particle-vibration coupling to reveal the underlying neutron orbitals outside $N=82$ isotones \cite{Ben-ah, Ben-Xe};  single-neutron states in $N=51$ nuclei \cite{DKS}; and a detailed study of the single-particle properties in Ni isotopes \cite{Ni1,Ni2}. 

This paper  focusses on a systematic study of hole states in the $N=82$ closed core. The low-lying structure of $N=81$ nuclei is largely based on configurations formed via core coupling with neutron holes in the shell between  $N=50$ and $N=82$ (see, for example, Reference~\cite{heyde}). This shell is composed of  $0g_{7/2}$, $1d_{5/2}$, $1d_{3/2}$, $2s_{1/2}$ and $0h_{11/2}$ single-particle orbitals, shown schematically in Figure~\ref{Sch}. The even-$Z$, $N=81$ isotopes that can be studied using stable beams and solid targets range from $^{137}_{56}$Ba to $^{143}_{62}$Sm. 

Light-ion nucleon-transfer reactions are a traditional tool with which to probe single-particle structure in nuclei and have been used for many years generating a wealth of information in the literature. However, systematic studies across chains of nuclei have been less common in the past and it can be difficult to use isolated studies to evaluate systematic trends as different experimental conditions and techniques have often been employed. In addition, the distorted-wave Born approximation (DWBA) calculations  required to extract spectroscopic information have been done with different computing codes and different choices of input parameters in different studies and were often limited by the  computation power available at the time, leading to the use of multifarious approximations. Indeed, the researcher trying to reassess experiments in the literature with modern reaction approaches is  stymied where the original absolute cross section data are not available in publications and only graphs of relative angular distributions or tables of spectroscopic factors are reported.

Here we describe a series of single-nucleon transfer experiments on stable solid $N=82$ targets, using a  magnetic spectrometer,  that have been used to determine the location of single-neutron hole strength in $N=81$ systems. These employ both the ($p$,$d$) and ($^{3}$He,$\alpha$) reactions  to ensure good momentum matching for low- and high-$\ell$ transfers,  respectively.  

There are several published works in the literature on hole strength, but systematic data across the solid stable $N=82$ targets using a consistent approach to both the experimental technique and the DWBA calculations with each reaction are not available. The ($p$,$d$) reaction has been studied previously on $^{138}$Ba, $^{140}$Ce, $^{142}$Nd and $^{144}$Sm targets, but with worse resolution than the current work \cite{JK1, JK2, Charm}. High-resolution measurements of the ($^{3}$He,$\alpha$) reaction were studied on  $^{140}$Ce, $^{142}$Nd and $^{144}$Sm targets in Ref. \cite{berrier}, which also reports measurements of the ($d$,$t$) reaction. However, the helium-induced reaction on a $^{138}$Ba target has not been studied before. In all this previous work, a zero-range approximation was used in the DWBA calculations and it was noted in several cases that there was sensitivity to some of the associated corrections \cite{JK1, JK2}. The calculations were also normalized by making assumptions about the single-particle purity of the $3/2^+$ ground states in each residual nucleus. Better approaches can now be employed to both DWBA calculations and the determination of their normalization. In addition to these studies, there are also a number of publications of reactions on isolated targets \cite{G,D,V,Y,F,K}. 

The current publication is organized in the following manner. Aspects of the experimental methodology will be discussed first, covering neutron removal with both ($p$,$d$) and ($^{3}$He,$\alpha$) reactions. The approach used to the DWBA calculations and normalization of the calculated cross sections follows, and the deduced single-neutron energies  will then be compared to a  simple model based on a two-body effective interaction between protons and neutrons.

\section{Experimental details}

Beams of 23-MeV protons and 34-MeV $^3$He ions were provided by the  tandem Van de Graaff accelerator at the A. W. Wright  Nuclear Structure Laboratory of Yale University. These beams were used to bombard targets of $^{138}$Ba, $^{140}$Ce, $^{142}$Nd and $^{144}$Sm. Momentum analysis of the ejectile ions was performed using the Yale Enge Split-Pole Spectrograph. At the focal plane, a multiwire gas proportional counter, backed by a plastic scintillator, was used to measure position, energy loss and residual energy of the ions passing through it. The ions were identified by combining information on  magnetic rigidity and energy-loss characteristics in the gas detector. The  beam dose was measured using a current integrator connected to a tantalum beam stop positioned behind the target. A +300~V bias was applied to both the target frame and beam stop to suppress electron sputtering.  Beam currents were typically in the range 50 to 100 enA for each beam species. A 1.5-mm thick silicon  detector was mounted at 30$^{\circ}$ to the beam axis to monitor target thickness, although the ratio of elastic scattering to beam current varied by less than 3\% on individual targets during the experiment.
 
 Given the  reactivity of the chemical elements used as targets, oxygen is an inevitable contaminant and, to avoid complicated vacuum transfer procedures, targets were manufactured by evaporation of isotopically-enriched oxide material  onto supporting carbon foils of thickness 20-40$\mu$gcm$^{-2}$. Reactions on oxygen and carbon did not overly complicate the analysis since the kinematic properties of ejectile ions from the contaminant reactions were sufficiently different from those of interest to be easily identified.
  
To allow the extraction of absolute cross sections, a calibration of the target thickness and spectrograph acceptance was necessary. The product of these two quantities was determined for each target by elastic scattering of 15-MeV $\alpha$ particles into the spectrometer at a laboratory angle of 20$^{\circ}$. Under these conditions, the cross section is expected to be within 0.5\% of that for Rutherford scattering.  The spectrometer entrance aperture was fixed throughout the experiment. The systematic uncertainty in cross sections determined this way was estimated to be around 5\%. Details of the four target foils are given in Table \ref{tab:targets}, where the  thicknesses given assume a nominal acceptance of 2.8~msr,  determined by previous calibrations using an $\alpha$ source at the target position \cite{Jason-pc}.

\begin{table}[h!]
\caption{\label{tab:targets} Details of the $N=82$ target foils.}
\begin{ruledtabular}
\begin{tabular}{cccc}
Target & Nominal Thickness & Isotopic \\
 &  $\mu$g~cm$^{-2}$ & enrichment \% \\
\hline\\
$^{138}$Ba & 101 & 99.8(1) \\
$^{140}$Ce & 144 & 99.9(1) \\
$^{142}$Nd &  150 & 99.0(1) \\
$^{144}$Sm &  42 & 93.8(1) \\
\end{tabular}
\end{ruledtabular}
\end{table}

\begin{figure}[htb]
\centering
\includegraphics[width=0.5\textwidth]{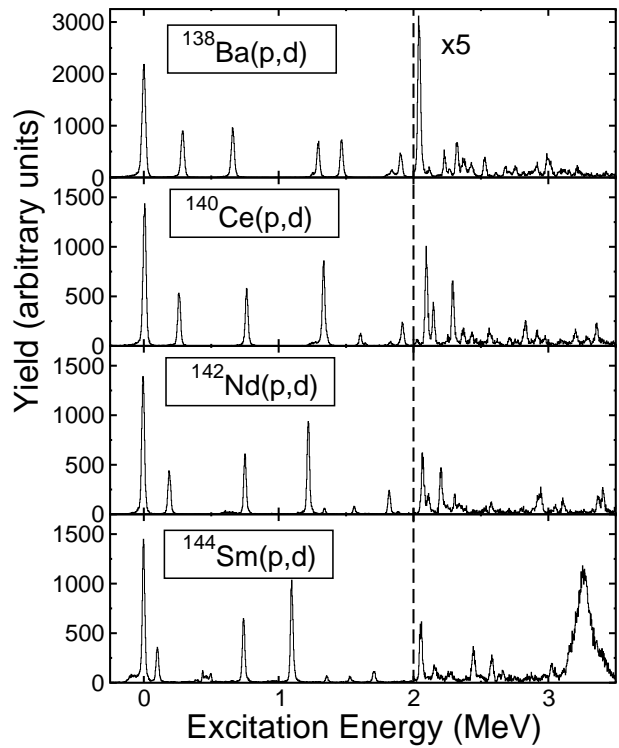}
\caption{\label{pd} Deuteron spectra from the ($p$,$d$) reaction on targets of $^{138}$Ba, $^{140}$Ce, $^{142}$Nd and $^{144}$Sm at an angle of 42$^\circ$, displayed in terms of the excitation energy of the residual nucleus. The portions of the  data to the right of the dotted line have been multiplied by a factor of five for clarity.}
\end{figure}

\begin{figure}[htb]
\centering
\includegraphics[width=0.5\textwidth]{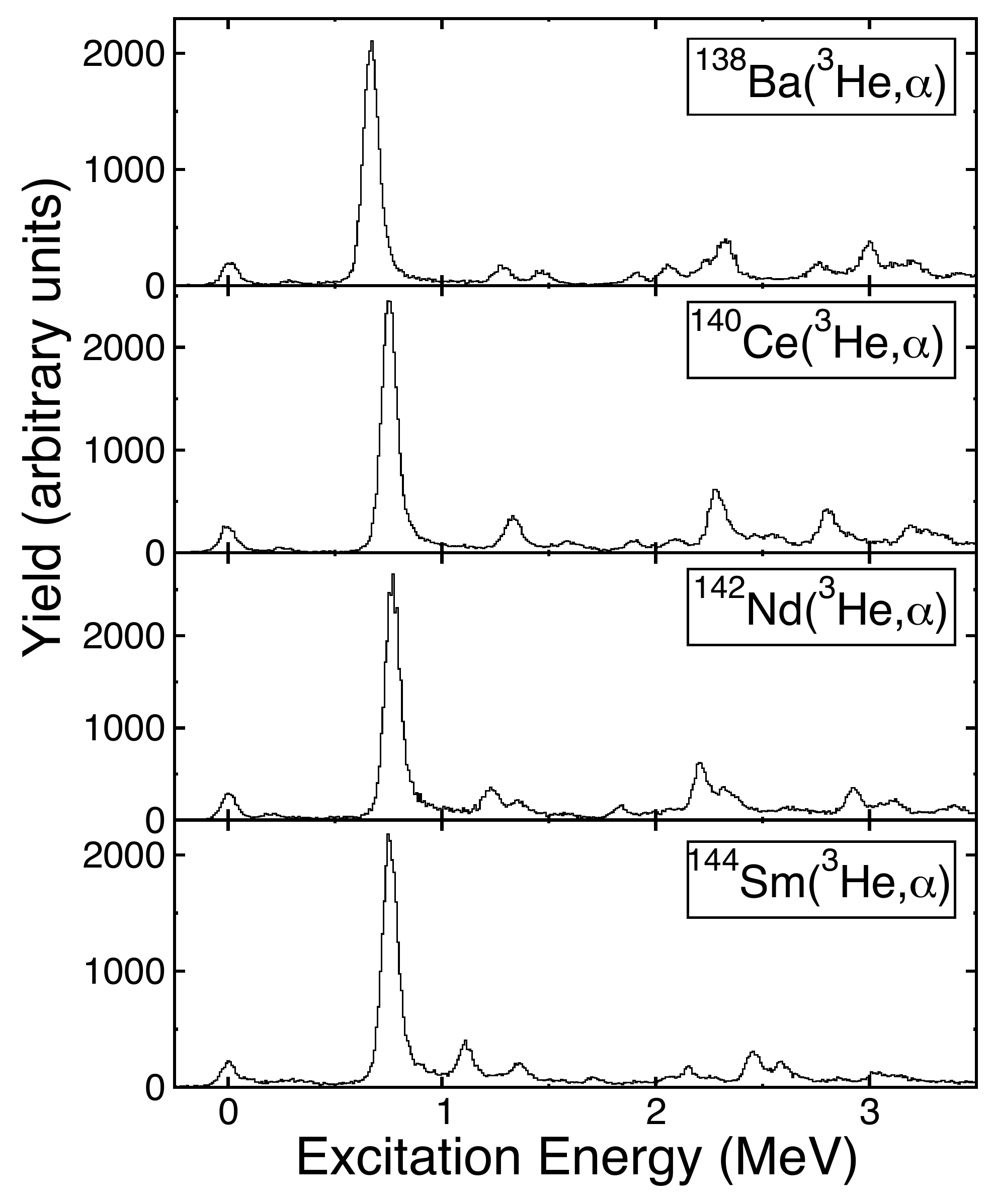}
\caption{\label{ha} $\alpha$-particle spectra from the ($^3$He,$\alpha$) reaction on targets of $^{138}$Ba, $^{140}$Ce, $^{142}$Nd and $^{144}$Sm at an angle of 15$^\circ$, displayed in terms of the excitation energy of the residual nucleus.}
\end{figure}

Representative focal-plane spectra for each target and reaction are shown in Figures~\ref{pd} and \ref{ha}. Comparison of the ($p$,$d$) and ($^3$He,$\alpha$) data in each case highlight the $\ell$ sensitivity of the reaction mechanism; for example, the $\ell=2$ transitions to the 3/2$^{+}$ ground states are visibly stronger in the ($p$,$d$) reactions than the ($^3$He,$\alpha$) reactions, whose spectra are dominated by the $\ell=5$ population of an excited 11/2$^{-}$ state at excitation energies  ranging from 661 to 754~keV across the residual nuclei. These spectra were calibrated using previously observed states, usefully summarized in References~\cite{Browne20072173, BURROWS2001623, Tuli2001277, TULI2001605}. The energy resolution was determined to be $\sim$25~keV for ($p$,$d$) data and $\sim$85~keV for ($^3$He,$\alpha$). Information on the excitation energies of known states, along with a width calibration determined from resolved states, were used to assist the analysis of unresolved peaks, especially in the ($^3$He,$\alpha$) spectra. Weak contaminant peaks resulting from the small quantities of $^{13}$C and $^{18}$O  present in the target foils were readily identifiable by their characteristic kinematic shift with angle, which also ensured that states of interest were affected by contaminant contributions at no more than one measurement angle. 

Data were collected at laboratory angles of 5$^{\circ}$, 20$^{\circ}$, 35$^{\circ}$ and 42$^{\circ}$ for the ($p$,$d$) reaction, chosen to be close to the first maxima of the expected angular distributions for \mbox{$\ell =0, 2, 4$} and 5 transitions, respectively. The distributions for the  ($^3$He,$\alpha$) reaction tend to be less distinct and more forward peaked, so data were only taken at 5$^{\circ}$ and 15$^{\circ}$. An additional angle of 10$^{\circ}$ was measured for the $^{138}$Ba target to assist assignments since the reaction had not been studied previously.

\begin{figure}[h!]
\centering
\includegraphics[width=0.35\textwidth]{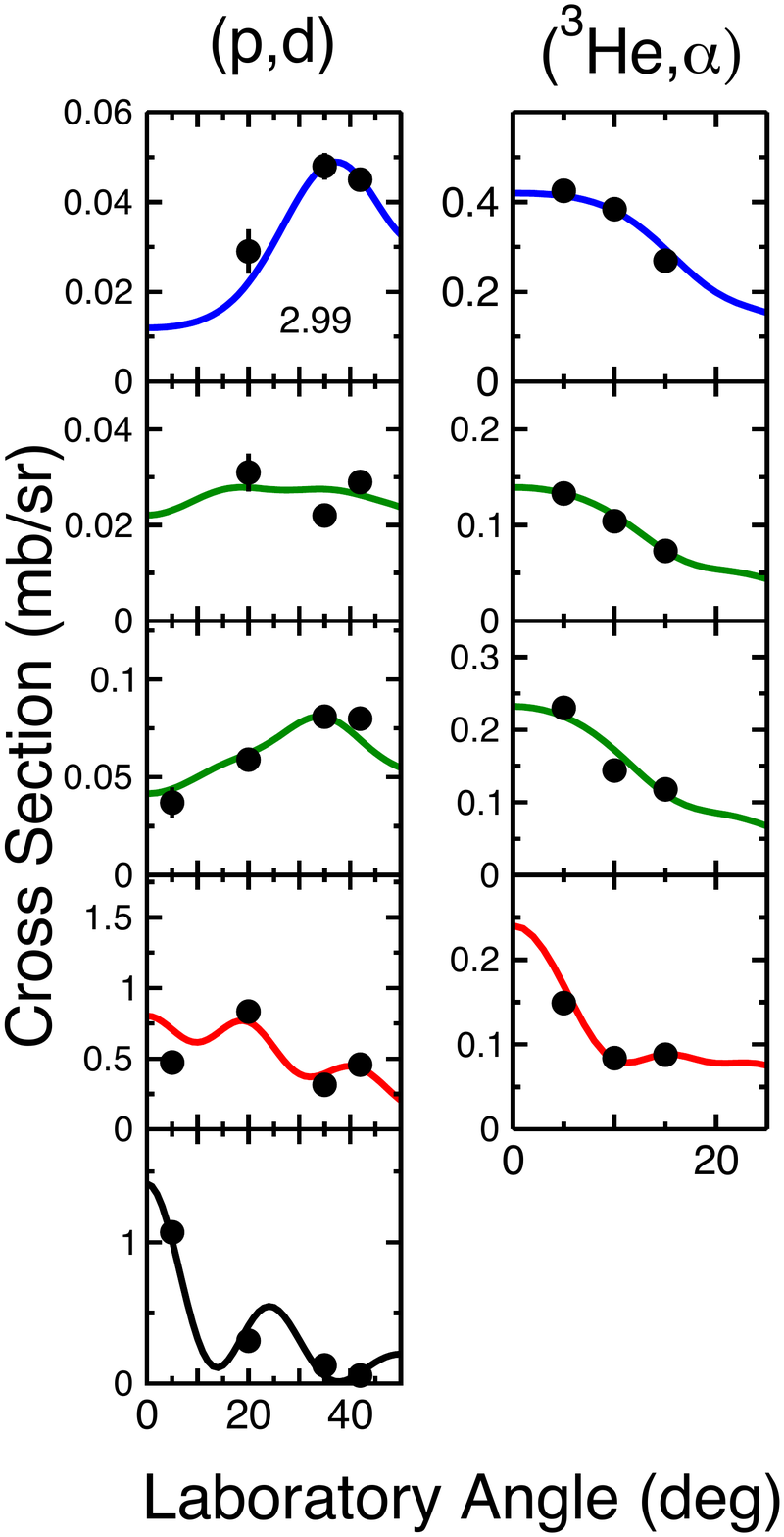}
\caption{\label{dist} Examples of angular distributions for the ($p$,$d$) and  ($^{3}$He,$\alpha$) reactions compared to the results of DWBA calculations discussed in Section III. The distributions are shown for states populated in $^{137}$Ba by $\ell=0$ (black), $\ell=2$ (red), $\ell=4$ (green) and $\ell=5$ (blue) transitions. Transitions with $\ell=0$ are not strongly populated in the ($^{3}$He,$\alpha$) reaction. The angular distributions are labeled with the excitation energy in the residual system in units of MeV.}
\end{figure}

\begin{figure}[h!]
\centering
\includegraphics[width=0.5\textwidth]{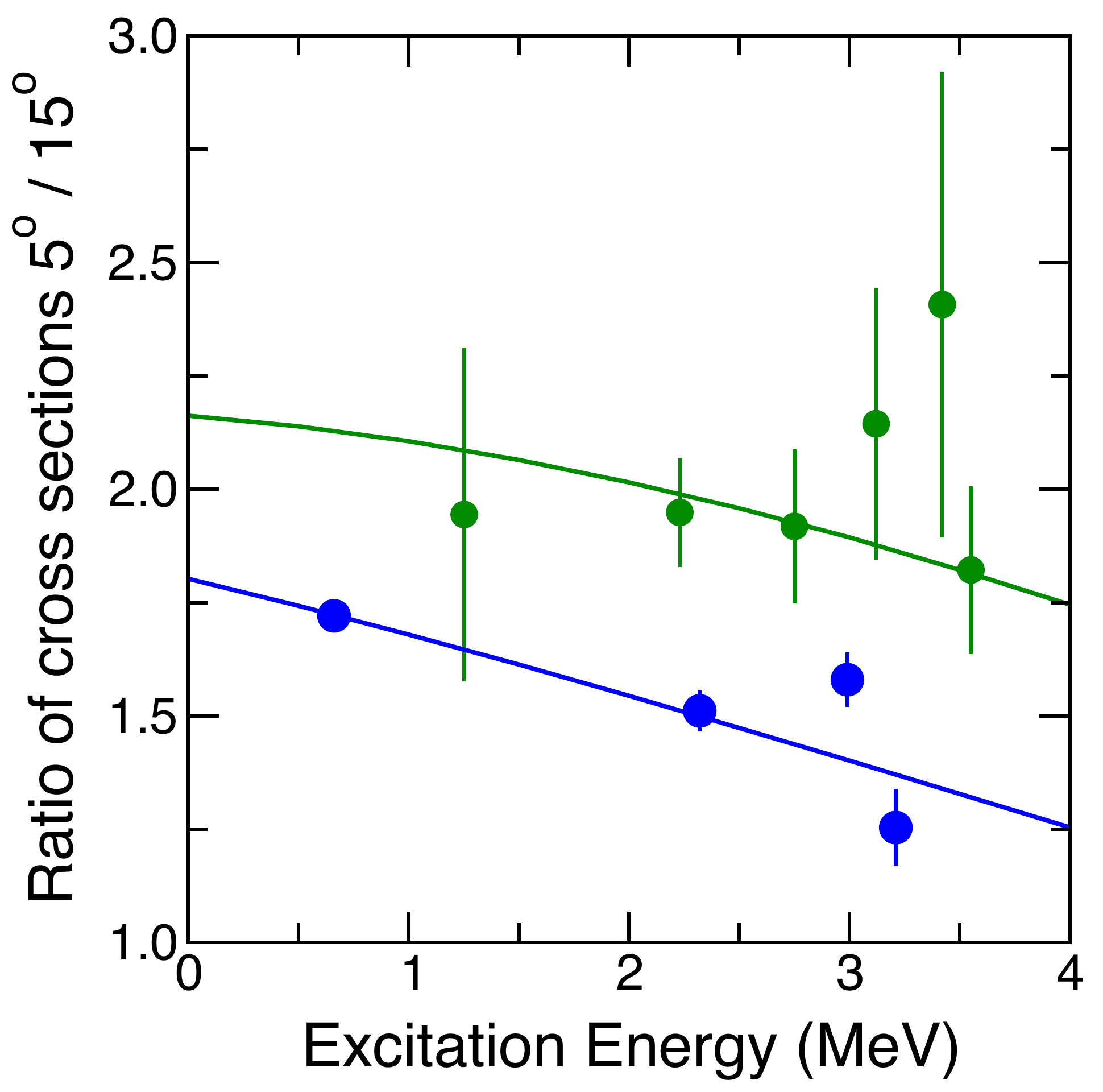}
\caption{\label{ha_slope} An example of the ratio of cross section at 5$^{\circ}$ and to that at 15$^{\circ}$ for the $(^{3}\text{He},\alpha)$ reaction, here shown for the population of states in $^{137}$Ba for $\ell=4$ (green) and $\ell=5$ (blue) as a function of excitation energy. The solid lines are the results of DWBA calculations  discussed in Section III.}
\end{figure}

For the majority of the states populated in the residual odd nuclei, angular-momentum quantum numbers have already been determined by a variety of different methods in the literature \cite{Browne20072173, BURROWS2001623, Tuli2001277, TULI2001605}. Previous assignments were checked using the following strategy. The angle of the first maxima of the angular distribution of the ($p$,$d$) reaction is generally indicative of the angular momentum transfer, so the shape of the ($p$,$d$) distribution was used in most cases to determine the $\ell$ values - some examples of angular distributions are shown in Figure~\ref{dist}. The angular distribution for $\ell=4$ transitions to states in the residual system were found to be increasingly flat  at higher excitation energies, behavior that is reproduced by DWBA calculations, but still distinct from those of $\ell=0,~2$ and 5 transitions. (Note that spectroscopic information for high-$\ell$ transfer is deduced from the ($^3$He,$\alpha$) reaction rather than from ($p$,$d$) cross sections, as discussed below). To confirm the assignments of high-$\ell$ transitions, the slopes of the ($^3$He,$\alpha$) angular distributions, in the form of the ratio of cross sections at 5$^{\circ}$ and 15$^{\circ}$, were also used, as illustrated in Fig.~\ref{ha_slope} for the $^{138}$Ba target. A comparison of the two differently-matched reactions has proved valuable in other work in differentiating between high-$\ell$ assignments (some examples can be found in References \cite{DKS,A=100, Ni2}); it was found to be less useful here in that respect, but did help to discriminate between high-$\ell$ and low-$\ell$ transitions. 

The ${\ell}$ values deduced from the current work for the three heaviest targets are generally consistent with the work on ($d$,$t$) and  ($^3$He,$\alpha$) reactions by Berrier {\it et al.} \cite{berrier}.  There is very good agreement for $^{141}$Nd. We note only minor discrepancies with Ref.~\cite{berrier} in $^{139}$Ce; strength at 2.910 and 3.352~MeV had previously each been found to carry {\it both} $\ell=2$ and $4$, but here no evidence for the presence of $\ell=4$ is found in the former and conversely, no evidence for $\ell=2$ in the latter. The population of the state at 2.018~MeV has been noted by several authors to have a non-standard distribution in neutron-removal reactions, which is confirmed here and no firm assignment could be made. The current work finds evidence for the presence of a tentative $\ell=0$ contribution at 2.556~MeV, along with the stronger $\ell=4$ transition. Spectroscopic factors for this doublet were determined on the basis that the ($p$,$d$) cross section at forward angles is due to the  $\ell=0$ strength and that this component does not contribute to the ($^3$He,$\alpha$) cross section, which was attributed entirely to $\ell=4$.

Assignments in $^{143}$Sm also agree well with Ref.~\cite{berrier}. However, at a beam energy of  23~MeV,  elastically-scattered protons have a lower kinetic energy and magnetic rigidity than deuterons arising from the population of the ground-state groups in the ($p$,$d$) reaction. Whilst the proton groups are fairly well separated from deuterons by energy-loss characteristics, a proton tail does contaminate the deuteron gating conditions, especially at  larger angles. This is the origin of the broad peak above 3~MeV in the $^{144}$Sm($p$,$d$) reaction in Figure~\ref{pd}. Similar groups in data on other targets lie higher in effective excitation energy than was studied here. Previous work has been performed  at higher energies \cite{G}, moving the elastic group to higher effective excitation energies, which circumvented this issue. The ($^3$He,$\alpha$) reaction does not suffer the same problem with elastic scattering, but  without the ($p$,$d$) data, assignments are more difficult. The two states at 3.13 and 3.23~MeV observed in the current work with the ($^3$He,$\alpha$) reaction are likely to be populated via high-$\ell$ transitions, but differentiation between $\ell=4$ and 5 has not been possible. For the later discussion, unobserved $\ell=5$ transitions would be a more critical issue; Ref.~\cite{berrier} observes no further $\ell=5$ population, whereas Ref.~\cite{G} isolates two higher-lying $\ell=5$ transitions. If the states at 3.13 and 3.23~MeV were $\ell=5$, it would shift the centroid of that strength in $^{143}$Sm by around 100~keV, which would not significantly alter the interpretation presented below.
 
In $^{137}$Ba, assignments up to 2~MeV are in agreement with those of previous ($p$,$d$) reactions \cite{JK2,Charm}. The $7/2^{+}$ peak at 1.252~MeV  in the current work, also observed by several other techniques \cite{Browne20072173}, has a $J^{\pi}$ assignment from $\gamma$-decay measurements following Coulomb excitation \cite{Ba_Coulex}. It was missed in both  previous ($p$,$d$) experiments, presumably masked by its more intense $\ell=2$ neighbour at 1.290~MeV. Ref.~\cite{JK2} also identified tentative assignments of the 7/2$^{+}$ state at 2.230~MeV and the 11/2$^{-}$ state at 2.320~MeV, which are confirmed here and supported by the ($^3$He,$\alpha$) data for the first time. The $\ell=4$ transitions also found in that work at 2.54 and 2.99~MeV have been revised here as $\ell=2$ and $\ell=5$, respectively. The former state is not observed strongly in the ($^3$He,$\alpha$) reaction, so the $\ell=4$ assignment of Ref.~\cite{JK2} is not confirmed. The latter state has  angular distributions in both reactions that are more consistent with $\ell=5$. The previous $\ell=4$ assignment in Ref.~\cite{JK2} may have been affected by the state at 3.03~MeV, which was unresolved from that at 2.99~MeV; the states were resolved, but no assignment was made, in Ref.~\cite{Charm}. In addition, 11 new assignments in $^{137}$Ba are made here, mainly $\ell=2$ states at excitation energies above 2.3~MeV.

The energies and $\ell$ assignments of all states observed are summarized in Table~\ref{SF}, along with spectroscopic factors determined using the procedures outlined below. Detailed data on cross sections are available as Supplemental Information \cite{supplemental}. The $J^\pi$ values listed in this table are taken from other measurements \cite{Browne20072173, BURROWS2001623, Tuli2001277, TULI2001605}; where $J^\pi$ assignments are not available, the subsequent analysis takes a model-dependent assumption that the strength is from the valence shell. However, in many cases, there is insufficient information to properly assign spin-parity to $\ell=2$ strength. 

\begin{figure*}
\centering
\includegraphics[scale=0.65]{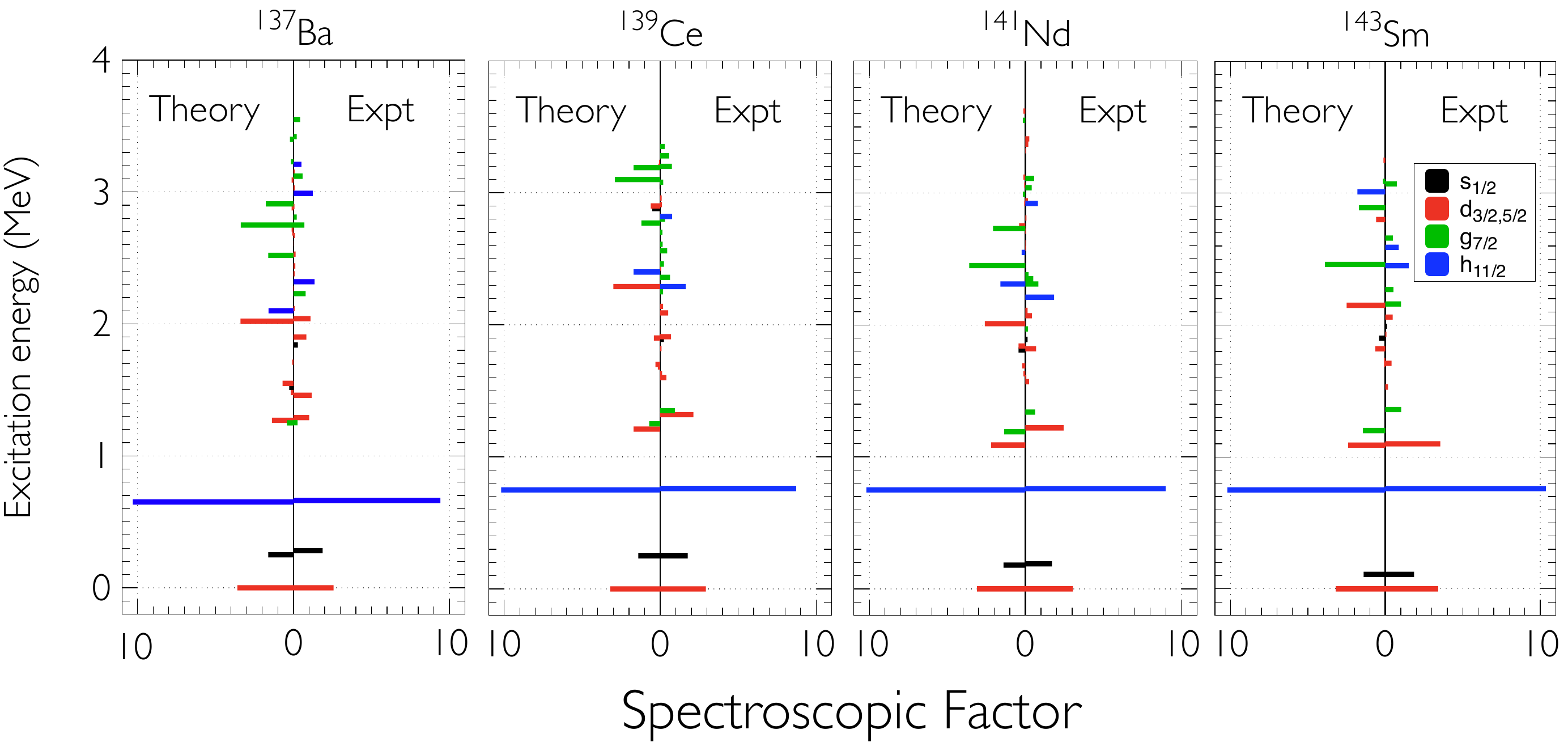}
\caption{\label{strength_distribution} Distribution of spectroscopic strength of states populated in ($p$,$d$) and $(^{3}\text{He},\alpha)$ reactions  for $\ell=0$ (black),  $\ell=2$ (red), $\ell=4$ (green) and $\ell=5$ (blue) transitions as a function of the excitation energy in the residual systems, compared to particle-vibration coupling calculations from Ref.~\cite{heyde}. The strength of individual states has been obtained from measured reaction cross sections using  procedures described in Section III.}
\end{figure*}

Although the extraction of single-particle strength using DWBA calculations is not discussed until the following section,  it is useful at this point to consider the general picture of the strength distributions in the residual nuclei, which is illustrated in Figure~\ref{strength_distribution}; the comparison with particle-vibration coupling calculations will be discussed later. The general pattern of behavior is similar to that revealed in neutron-removal reactions on $^{134,136}$Ba \cite{Swzec} and $^{128,130}$Te \cite{Kay}. The ground state in each case is a $3/2^{+}$ state carrying a significant fraction of the expected $d_{3/2}$ strength, increasing with $Z$ from around 64\% in $^{137}$Ba to 85\% in $^{143}$Sm. Older studies have made the assumption that this state carries all of the $d_{3/2}$ strength \cite{JK1,JK2, Charm}. At a few 100~keV in excitation energy, there is a $1/2^{+}$ state with significant $s_{1/2}$ strength (90\% on average and not varying significantly across the isotopes). Beyond that lies a strong $11/2^-$ state with around 80\% of the expected $h_{11/2}$ strength. These correspond to the three low-lying strong peaks that can be seen in the ($p$,$d$) spectra (see Fig.~\ref{pd}) and the population of the $11/2^-$ state dominates the $(^{3}\text{He},\alpha)$ spectra (see Fig.~\ref{ha}). At higher excitation energies, there is a second strong $\ell=2$ transition above 1~MeV, obvious in the ($p$,$d$) reactions on $^{140}$Ce, $^{142}$Nd and $^{144}$Sm targets, which has been given a $5/2^{+}$ assignment in other work, carrying between 35 and 50\% of the $d_{5/2}$ strength. In $^{137}$Ba, the corresponding state has a lower strength and an additional, relatively strong $3/2^+$ state occurs just above in excitation energy. 

Above $\sim$1.8~MeV in each residual nucleus, there are numerous small fragments of strength, which appear to be dominated by $\ell=2$ and $\ell=4$ strength, with a few even weaker isolated $\ell=0$ and $\ell=5$ transitions. It therefore appears that  most of the strength associated with the $s_{1/2}$, $d_{3/2}$ and $h_{11/2}$ orbitals are generally contained in a low-lying state with low levels of fragmentation. The low-lying $\ell=4$ state apparent around 1.2~MeV in Sm, Nd and Ce final nuclei only carries only around 10\% of the $g_{7/2}$ strength, the rest is dispersed in small fragments at high excitation energies with a significant proportion lying at higher excitation energies than studied here;  this 10\% fragment does not appear in $^{137}$Ba. Across all the residual nuclei the deeper-lying $d_{5/2}$ and $g_{7/2}$ hole strengths are significantly fragmented over many states extending to high excitation energies. 

\begin{turnpage}
\squeezetable
\begin{table*}[h!]
\caption{\label{SF} Summary of states populated in neutron-removal reactions on $N=82$ targets, including excitation energy $E$, orbital angular momentum transfer $\ell$, spin-parity $J^{\pi}$, and normalised spectroscopic factor $C^2S$.  Excitation energies are given in MeV and are estimated to carry an uncertainty of $\sim$5~keV, rising to $\sim$10~keV in the case of $^{137}$Ba at higher excitation energies. The spectroscopic factors are deduced from the ($p$,$d$) reaction for $\ell=$0 and 2 transfers and from the $(^{3}\text{He},\alpha)$ reaction for $\ell=$4 and 5, and have been normalized using the method described in the text. The errors in the normalised values are typically 5\% due to variation of DWBA with different input parameters, but for weaker transitions these rise where statistical errors become more significant (more information is available in the Supplemental Material \cite{supplemental}). $J^{\pi}$ are taken from the literature \cite{Browne20072173, BURROWS2001623,  Tuli2001277, TULI2001605}; where a $J^\pi$ value is not listed, a model-dependent assumption was made that the single-particle orbitals is in the valence shell.}
\begin{ruledtabular}
\begin{tabular}{*{20}{lcccclcccclcccclcccc}}
\\
 \multicolumn{4}{c}{$^{137}$Ba}&&\multicolumn{4}{c}{$^{139}$Ce}&&\multicolumn{4}{c}{$^{141}$Nd}&&\multicolumn{4}{c}{$^{143}$Sm}\\
 \\
 \hline
 \\
E&$\ell$&$J^{\pi}$&$C^2S$&~~~~~~~~&E&$\ell$&$J^\pi$&$C^2S$&~~~~~~~~&E&$\ell$&$J^\pi$&$C^2S$&~~~~~~~~&E&$\ell$&$J^\pi$&$C^2S$\\
\\
\hline
\\
Ê0.000	&	2	&	3/2$^+$	&	2.56	&	&	0.000	&	2	&	3/2$^+$	&	2.92	&	&	0.000	&	2	&	3/2$^+$	&	3.04	&	&	0.000	&	2	&	3/2$^+$	&	3.40	 \\
Ê0.281	&	0	&	1/2$^+$	&	1.86	&	&	0.252	&	0	&	1/2$^+$	&	1.77	&	&	0.192	&	0	&	1/2$^+$	&	1.69	&	&	0.110	&	0	&	1/2$^+$	&	1.84	 \\
Ê0.662	&	5	&	11/2$^-$	&	9.42	&	&	0.755	&	5	&	11/2$^-$	&	8.72	&	&	0.759	&	5	&	11/2$^-$	&	8.99	&	&	0.758	&	5	&	11/2$^-$	&	10.36	 \\
Ê1.252	&	4	&	7/2$^+$	&	0.26	&	&	1.321	&	2	&	5/2$^+$	&	2.12	&	&	1.222	&	2	&	5/2$^+$	&	2.44	&	&	1.100	&	2	&	5/2$^+$	&	3.54	 \\
Ê1.290	&	2	&	5/2$^+$	&	1.01	&	&	1.347	&	4	&	7/2$^+$	&	0.94	&	&	1.343	&	4	&	7/2$^+$	&	0.62	&	&	1.362	&	4	&	7/2$^+$	&	1.02	 \\
Ê1.460	&	2	&	3/2$^+$	&	1.17	&	&	1.598	&	2	&	(3/2)$^+$	&	0.40	&	&	1.565	&	2	&	(3/2)$^+$	&	0.23	&	&	1.533	&	2	&	(5/2)$^+$	&	0.17	 \\
Ê1.840	&	0	&	1/2$^+$	&	0.28	&	&	1.632	&	2	&	3/2$^+$	&	0.12	&	&	1.597	&	2	&	(3/2,5/2)$^+$	&	0.06	&	&	1.708	&	2	&	(3/2)$^+$	&	0.40	 \\
Ê1.900	&	2	&	3/2$^+$	&	0.83	&	&	1.823	&	2	&	5/2$^+$	&	0.09	&	&	1.822	&	2	&		&	0.69	&	&	1.930	&	2	&		&	0.07	 \\
Ê2.040	&	2	&	(5/2)$^+$	&	1.08&	&	1.889	&	0	&	1/2$^+$	&	0.24	&	&	1.888	&	0	&		&	0.14	&	&	1.990	&	0	&		&	0.11	 \\
Ê2.117	&	2	&		&	0.07	&	&	1.911	&	2	&	(3/2)$^+$	&	0.69	&	&	1.968	&	4	&	7/2$^+$	&	0.17	&	&	2.064	&	2	&		&	0.47	 \\
Ê2.230	&	4	&	7/2$^+$	&	0.78	&	&	2.018	&		&		&		&	&	2.070	&	2	&	(3/2$^+$,5/2$^+$)	&	0.41	&	&	2.161	&	4	&	7/2$^+$	&	1.01	 \\
Ê2.271	&	2	&	(3/2$^+$,5/2)	&	0.04	&	&	2.090	&	2	&		&	0.51	&	&	2.111	&	2	&	3/2$^+$,5/2$^+$	&	0.14	&	&	2.274	&	4	&	7/2$^+$	&	0.52	 \\
Ê2.32	&	5	&		&	1.35	&	&	2.143	&	2	&		&	0.19	&	&	2.180	&	0	&		&	0.05	&	&	2.450	&	5	&		&	1.52	 \\
Ê2.38	&	2	&		&	0.07	&	&	2.251	&	(4)	&	(7/2$^+$)	&	0.19	&	&	2.208	&	5	&	(11/2)$^-$	&	1.84	&	&	2.586	&	5	&		&	0.87 \\
Ê2.44	&	2	&		&	0.12	&	&	2.286	&	5	&	11/2$^-$	&	1.63	&	&	2.31	&	4	&	7/2$^+$,(9/2$^+$)	&	0.83	&	&	2.662	&	4	&		&	0.48	 \\
Ê2.53	&	2	&		&	0.14&	&	2.362	&	4	&		&	0.63	&	&	2.349	&	4	&		&	0.50				&	&	3.05	&(4)		&		&0.64		 \\
Ê2.61	&	(2)	&		&	0.02	&	&	2.426	&	2	&		&	0.06	&	&	2.384	&	4	&	7/2$^+$	&	0.20		&	&	3.13	&		&		&		 \\
Ê2.67	&	2	&		&	0.09	&	&	2.455	&	(4)	&		&	0.24	&	&	2.512	&		&		&		&				&	3.23	&		&		&		 \\
Ê2.75	&	4	&		&	0.70	&	&	2.556	&	4 \& (0)	&		&	0.45 \& 0.04	&	&	2.581	&	(2)	&		&	0.05	&	&		&		&		&		 \\
Ê2.81	&	(4)	&		&	0.21	&	&	2.610	&	(4)	&		&	0.16	&	&	2.616	&	(2)	&		&	0.02	&	&		&		&		&		 \\
Ê2.89	&	2	&		&	0.06	&	&	2.701	&	(4)	&		&	0.14	&	&	2.705	&	(2)	&		&	0.05&	&		&		&		&		 \\
Ê2.99	&	5	&		&	1.24	&	&	2.800	&	4	&	7/2$^+$	&	0.31	&	&	2.809	&	(2)	&		&	0.07&	&		&		&		&		 \\
3.03	&	2	&		&	0.09	&	&	2.822	&	5	&	9/2$^-$,11/2$^-$	&	0.76	&	&	2.915	&	5	&		&	0.80	&	&		&		&		&		 \\
Ê3.12	&	4	&		&	0.58	&	&	2.910	&	2	&		&	0.11&	&	2.939	&	2	&		&	0.16	&	&		&		&		&		 \\
Ê3.15	&	(2)	&		&	0.07	&	&	2.964	&	2	&		&	0.10&	&	3.042	&	4	&		&	0.40	&	&		&		&		&		 \\
Ê3.21	&	5	&		&	0.51	&	&	3.082	&	(4)	&		&	0.20	&	&	3.112	&	4	&		&	0.56	&	&		&		&		&		 \\
Ê3.42	&	4	&		&	0.21	&	&	3.196	&	4	&		&	0.75	&	&	3.315	&	(2)	&		&	0.04&	&		&		&		&		 \\
Ê3.55	&	4	&		&	0.43	&	&	3.282	&	4	&		&	0.58	&	&	3.369	&	2	&		&	0.19	&	&		&		&		&		 \\
	&		&		&		&	&	3.352	&	4	&		&	0.30	&	&	3.407	&	2	&		&	0.25	&	&		&		&		&		 \\
Ê\end{tabular}
\end{ruledtabular}
\end{table*}
\end{turnpage}


\section{DWBA  and normalization}

Spectroscopic factors were determined from the measured cross sections by comparison with the results of calculations using the distorted-wave Born approximation with the finite-range code PTOLEMY \cite{ptolemy}.  The approach taken here is same procedure adopted in a recent global analysis of quenching of spectroscopic strength \cite{quench}, which has also been used in a number of recent studies, for example Refs.~\cite{A=100, szwec, Jonathan}. The choices for potentials associated with the optical models describing the initial and final reaction channels, and those associated with the neutron bound states in the light and heavy cores, are  the same as those used previously, with one minor exception, and are summarized below.  

The incoming and outgoing partial waves were described using the  global optical potentials for protons \cite{koning}, deuterons \cite{perey}, and helions \cite{pang}. The  deuteron potential used here gave a better reproduction of the angular distributions than more recent global potentials \cite{ann} that we have employed in previous cases. The potential of Ref.~\cite{perey} had been used as the starting point in the search for new parameters to extend the potential to wider energy range in Ref.~\cite{ann}, but the current deuteron energies are within those used in the former potential.  A fixed $\alpha$-particle potential determined from the $A=90$ region was used \cite{Bassani}.

Recent microscopic calculations were used as the source for the internal wave functions of the light ions in the reactions. For the deuteron, form factors determined using the Argonne $v_{18}$ potential were used \cite{AV18} and those for the $\alpha$ particle and $^3$He ions were taken from recent Green's function Monte-Carlo calculations \cite{GMC}. 

The wave functions of the transferred neutron in the heavy bound state were generated using a Woods-Saxon potential with a depth adjusted to match the measured binding energy. This used  fixed geometric parameters: radius parameter $r_0$=1.28~fm and diffuseness $a=0.65$~fm. The derivative of a Woods-Saxon potential with radius  $r_{so}$=1.10~fm, diffuseness $a_{so}=0.65$~fm and depth $V_{so}$=6~MeV was used to model the spin-orbit component.

The approximations involved in the DWBA approach are best satisfied where there is a large probability of a direct reaction mechanism. Spectroscopic factors are therefore extracted using experimental cross sections measured as close as possible to the angle of the first maximum of the angular distribution of the most appropriately matched reaction. The ($p$,$d$) reaction was used to determine the spectroscopic strength for $\ell=0$ and $2$ from data at $5^{\circ}$ and $20^{\circ}$, respectively, whereas that for $\ell=4$ and $5$ was extracted from the $(^{3}\text{He},\alpha)$ reaction at $5^{\circ}$.

The DWBA calculations carry an overall uncertainty in absolute normalization. Consistent results have been obtained by adopting systematic approaches (for example, Ref.~\cite{Ni1, Ni2})  using the Macfarlane-French sum rules \cite{macf} which associate the summed spectroscopic strengths to the occupancies and vacancies of single-nucleon orbitals. If a normalization factor is chosen such that the total observed strength is equal to the full single-particle value, the degree to which that   factor deviates from unity is related to  quenching of single-particle strength. Such quenching has been observed in other reactions, such as ($e,e^{\prime}$p)  \cite{eep1, eep2}, where the total low-lying strength accounts for approximately half that expected by the independent-particle model. A recent large-scale analysis of transfer data has found normalization factors that are quantitatively consistent with previous studies of such quenching \cite{quench} and here we follow the same procedure.

The total spectroscopic strength was required to reproduce the number of expected neutrons in the corresponding orbital in the target nucleus.  On the assumption of the closed neutron shell at $N=82$, this  corresponds to the degeneracy of the orbital. This assumption can be tested by probing the vacancy of the orbitals below the shell closure by looking for population of the relevant $\ell$ transfer in ($d$,$p$) reactions on $N=82$ targets. Several such studies exist in the literature, but evidence for population of orbitals with the quantum numbers of the nominally-filled neutron orbitals is sparse and any such states are populated very weakly. As examples, Ref.~\cite{Park} observes an $\ell=0$ transition at 3.351~MeV and three tentative $\ell=2$ transitions above 2.2~MeV, with strengths of around 1\% in $^{141}$Ce. Ref.~\cite{Veefkind} reports an  $\ell=0$ transition at 1.616~MeV in $^{143}$Nd with a similar intensity. Such weak transitions are also likely to be subject to higher contributions from indirect processes.  There appears to be no evidence for the relevant $\ell$ transfer in $^{139}$Ba or $^{145}$Sm. The assumption of a closed shell looks reasonable, at least compared to other uncertainties.

Initially normalization was performed separately for each $\ell$ value in the appropriately matched reaction and the results are shown in in Table~\ref{N-Table}.

\begin{table}[H]
\caption{\label{N-Table} Normalization factors for DWBA calculations with the associated mean and standard deviation across the four targets studied. Asterisks indicate cases that are affected by significant unobserved strength.\\}
\begin{ruledtabular}
\begin{tabular}{r c c c r}
 & \multicolumn{2}{c}{($p,d$)}&\multicolumn{2}{c}{($^3$He,$\alpha$)}\\
 & $\ell=0$&$\ell=2^*$ &$\ell=4^*$  &$\ell=5$\\
\hline\\
$^{138}$Ba	& 0.58& 	0.40&	0.22&	0.58\\
$^{140}$Ce	& 0.55&  0.40&	0.40&	0.52\\
$^{142}$Nd	& 0.51&  0.42&	0.23&	0.54\\
$^{144}$Sm	& 0.53&  0.44&	0.31&	0.59\\
\hline
Mean 		& 0.54	&	0.41&	0.27&	0.56\\
St Dev & 0.03 & 0.02&0.06&0.04\\
 \end{tabular}
 \end{ruledtabular}
 \end{table}

The mean normalization factors for the $\ell=0$ and $\ell=5$ are 0.54 and 0.56, respectively, with a variation of around 0.03 across the targets. These values compare  favourably with a recent systematic analysis of transfer data on targets from $^{16}$O to $^{208}$Pb for a variety of different proton and neutron transfer reactions over a range of $\ell$ values, which deduced a quenching with respect to independent-particle models of 0.55 \cite{quench}. The mean quenching factors deduced in that work for low $\ell$ transitions in ($d$,$p$) and ($p$,$d$) reactions was 0.53; the excellent correspondence with the current normalization for $\ell=0$ is particularly encouraging. It relieves a potential concern that, given measurements at 0$^{\circ}$ are not possible, $\ell=0$ spectroscopic factors cannot be obtained as close to the first maximum of the angular distribution as other $\ell$ values and, by necessity, are extracted in a region of a rather strongly sloping angular distribution.

However, the average values for $\ell=2$ and $\ell=4$, at 0.41 and 0.27, respectively, are significantly lower. This suggests that the experiment is missing some of the low-lying strength associated with the corresponding orbitals. This finding is not inconsistent with the observed distribution of high-lying, dispersed and fragmented strength for $\ell=2$ and 4 (see Fig.~\ref{strength_distribution}) where the risk of missing strength is high, either in the form of transitions lying outside the measured excitation range or in the form of  small unresolved fragments of strength in the measured spectra. We therefore adopt the values of 0.54 and 0.56 for the DWBA normalizations for the ($p,d$) and ($^3$He,$\alpha$) reactions, respectively.

The choice of potentials used in the DWBA calculation has a significant effect on the absolute magnitude of the raw unnormalised spectroscopic factors; calculations were repeated with a number of other physically reasonable  potentials and a variation of $\sim$20\% in the calculated absolute cross sections was found. {\it Normalised} spectroscopic factors, determined using the procedures outlined above, are far less sensitive to choices of optical models and were found to vary by around $\sim$5\%. The influence of multi-step processes  is expected to be similar to that estimated in other  analyses \cite{Ni2, A=100} and are a less significant effect.

There is a small complication that arises for neutron-removal (and proton-adding) reactions associated with isospin effects. In these  reactions, the transfer results in the population of states with both isospin couplings,  $T\pm1/2$ where $T$ is the target isospin. The states corresponding to the higher isospin coupling  $T_>$ lie at excitation energies higher than those accessed here experimentally. In principle, the Macfarlane and French sum rules used  in the normalization procedure for neutron-removal reactions need to include the $T_>$ strength. This can be done on the basis of isospin symmetry, using spectroscopic factors $C^2S$ for analogous states in proton-removal reactions and applying the appropriate isospin Clebsch-Gordan coefficients to deduce the spectroscopic factor associated with the higher isospin \cite{Schiffer}. 

\begin{figure}[h!]
\centering
\includegraphics[width=0.5\textwidth]{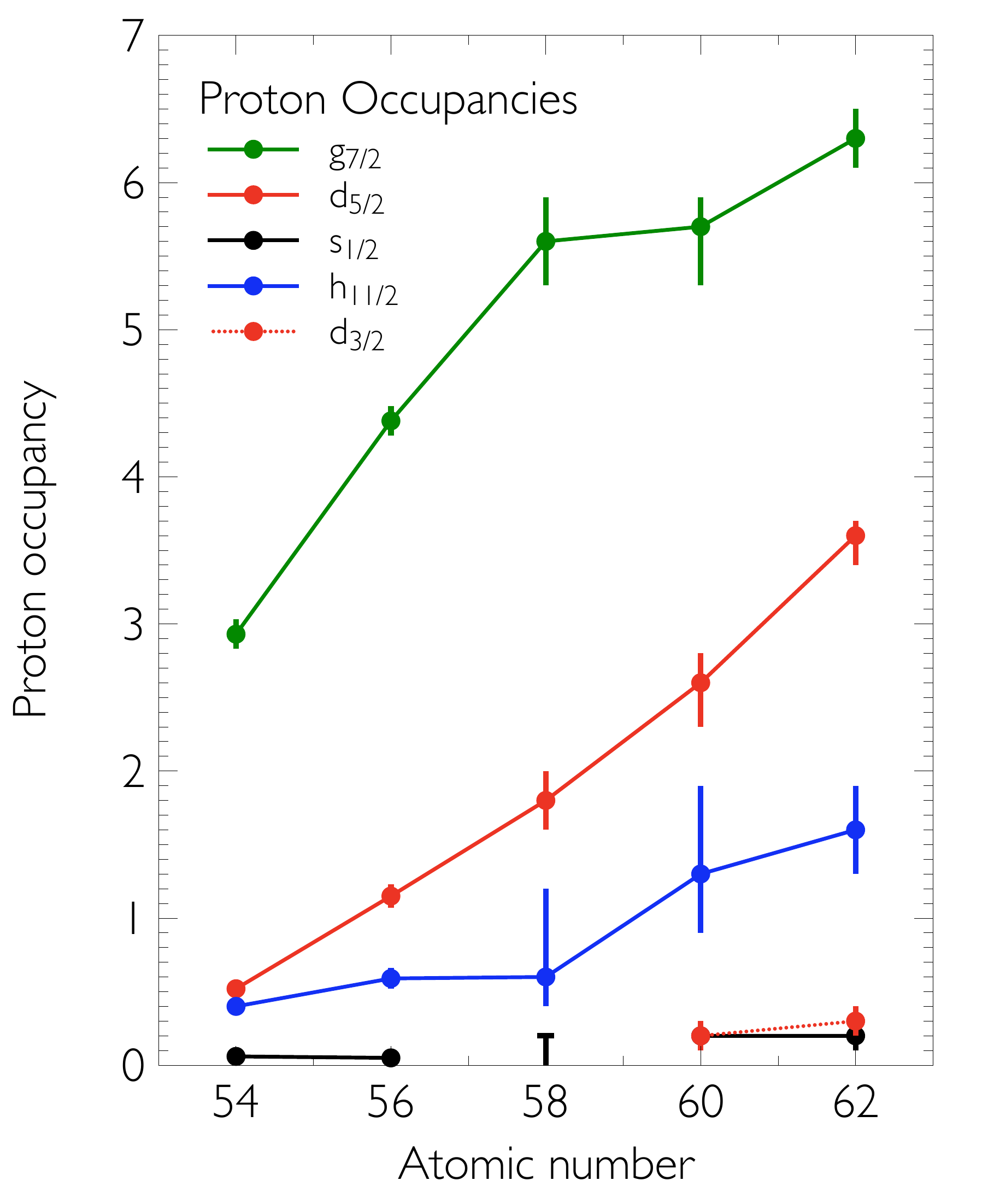}
\caption{\label{occ} Occupancy of single-proton orbitals in $N=82$ nuclei as a function of proton number, taken from  from Ref.~\cite{Wild} for Ce, Nd and Sm and Ref.~\cite{Jonathan} for Xe and Ba. No proton strength was observed for the  $s_{1/2}$ orbital in Ref.\cite{Wild} for Ce and an upper limit of 0.2 was placed on the associated occupancy.}
\end{figure}

The nuclei studied here are near the beginning of the $Z=50-82$ shell and protons are known to occupy mainly the $g_{7/2}$ and $d_{5/2}$ orbitals \cite{Wild}; the spectroscopic factors for proton removal from the $\ell=0$ and 5 orbitals relevant for the normalisation are consequently small (see Figure~\ref{occ}). Moreover, the ratios of isospin Clebsch-Gordan coefficients that are required to convert these into the spectroscopic factors for the higher isospin states in neutron removal are also small. The overall correction for the non-observation of the upper isospin component is less than a 1\% effect for these orbitals and is smaller than other uncertainties. The correction has therefore been neglected in the normalization procedure here. Larger corrections would apply to the summed strengths for $g_{7/2}$ and $d_{5/2}$, which have significant population of protons and large proton removal strengths, but these are not used to determine the normalization.

\section{Discussion}

\begin{figure}[h!]
\centering
\includegraphics[width=0.5\textwidth]{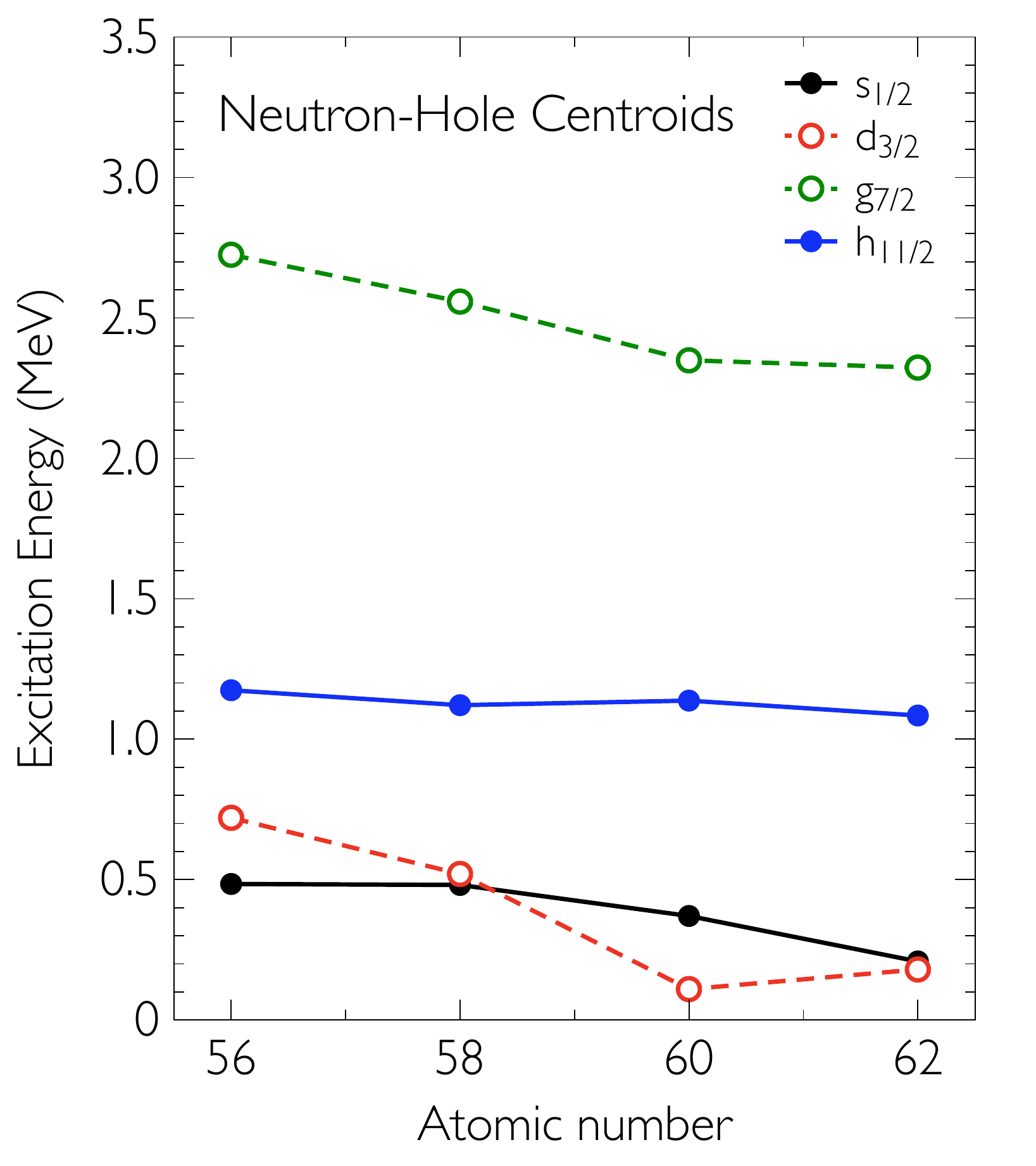}
\caption{\label{centroids} Variation in the excitation energy of the centroid of {\it observed} single-particle strength for the $T_<$ component as a function of proton number. Statistical errors are of the order $\sim$10~keV.  The open circles and dotted lines indicate instances where the full single-particle strength has not been observed. The centroid for the $d_{3/2}$ orbital uses states that have a $3/2^+$ spin-parity in the literature. The data for the $g_{7/2}$ orbital suffers from significant unobserved strength outside of the excitation-energy range measured and the true single-particle centroid will lie significantly higher than the observed centroid (see text for details).}
\end{figure}

Spectroscopic factors, extracted using the procedure outlined in the previous section, were used to determine the centroids of observed single-neutron hole strengths for the $T_<$ isospin components. These centroids and the associated summed strength are summarized in Table~\ref{C-Table} and shown as a function of atomic number in Figure~\ref{centroids}. 

In some previous studies, it has been assumed that the  3/2$^+$ ground state exhausted the $d_{3/2}$ strength, but here it is found that the associated spectroscopic factor increases from $^{137}$Ba to $^{143}$Sm. In addition to the total $\ell=2$ strength, Table~\ref{C-Table} also shows values associated with $\ell=2$ transitions populating states with a firm or tentative 3/2$^+$ spin assignment and the centroid of these are shown in Fig.~\ref{centroids}. The associated summed strengths are not as consistent across the isotopes as for the other $\ell$ values, indicating that in some cases there is missing $d_{3/2}$ strength and in others that there are likely some mis-assignments of $j$ values. The remaining $\ell=2$ strength is likely attributable to the $d_{5/2}$ orbital, but it varies between 50\% and 76\% of the full strength across the isotopes. Fragmentation is high and a significant portion of the strength lies at excitation energies higher than measured here.

In the case of the $g_{7/2}$ strength, there is significant missing strength and the current work only observed between 40 and 61\%, depending on the isotope. The true single-particle centroid lies higher than the observed centroid quoted in Table~\ref{C-Table}; we estimate that the true centroid lies {\it at least} 450, 350, 700 and 600~keV higher in energy than the observed centroids in $^{137}$Ba, $^{139}$Ce, $^{141}$Nd and $^{143}$Sm, respectively, and because of this large uncertainty, we make no further discussion of $\ell=4$ strength here.

 In the cases where most of the low-lying strength has been captured ($\ell=0$ and 5), the centroid across both $T_<$ and $T_>$ isospin components would reflect the underlying single-neutron energy. As discussed above, only the $T_<$ strength is observed in the current work. The location and strength of the $T_>$ component were estimated using Coulomb displacement energies and data from proton-removal reactions \cite{Wild} using isospin symmetry. It was found that the difference between the full centroid and that for the $T_<$ component of the $\ell=0$ and 5 strength increases with $Z$ from around 20 to 90~keV across the isotopes. This is relatively small since the associated orbitals have low proton occupancy. The correction is much larger for $\ell=2$ and 4 strength, but these are the same orbitals where significant strength remains unobserved in the current experiment and the interpretation of the measured centroids is difficult. We therefore use the variation in the measured centroids of $\ell=0$ and 5 strength as an estimate for the changes in the underlying single-neutron energies across the isotones studied. 
 
Changes in orbital energies across chains of nuclides have been interpreted in terms of the effect of valence proton-neutron interactions as the nucleon number varies. Here we follow the approach of Reference~\cite{Otsuka2} where changes in the effective single-neutron energies were compared to calculations using a two-body central plus tensor force between  neutrons and valence protons, taking information on proton occupancy from proton-transfer experiments in the literature.

\begin{table*}
\caption{\label{C-Table} Observed summed hole strengths and the associated centroid excitation energies for the $T_<$ components.  The summed strength is deduced from spectroscopic factors that were normalized using the method described in the text. The errors quoted on the summed strength are on the basis of the variations due to choices of potentials in the DWBA (see text for details). The errors on the centroid in the table are statistical. Values are given for the sum of $d_{3/2}$ and $d_{5/2}$ orbitals deduced for the $\ell=2$ transitions and also separately for states populated by $\ell=2$ transitions with a spin-3/2 assignment in the literature. Asterisks indicate cases that are affected by significant unobserved strength, which gives rise to a significant systematic uncertainty in the true single-particle centroid.\\}
\begin{ruledtabular}
\begin{tabular}{r c c c c c c c c c r}
 Orbital & \multicolumn{5}{c}{Summed Strength}&~~~& \multicolumn{4}{c}{Centroid Energy (MeV)}\\
  & $^{137}$Ba & $^{139}$Ce & $^{141}$Nd & $^{143}$Sm&Expected &&$^{137}$Ba & $^{139}$Ce & $^{141}$Nd & $^{143}$Sm\\
\hline
$s_{1/2}$	& 2.1(1)&2.0(1) &1.87(9) &1.9(1) & 2& & 0.48(1)&0.48(2)&0.37(1)&0.21(1)\\
$d^*$		& {7.4(4)}&{ 7.3(4) }&{7.8(4) }&{ 8.0(4) }& 10		& &{ 1.19(2)}&{ 1.01(2)}&{ 1.07(3)}&{ 0.74(3)}\\
$d_{3/2}$		& 4.6(2)&  4.1(2)&  3.26(16)& 3.8(2)& 4		&  & 0.72(2)& 0.52(2)& 0.11(2)& 0.18(2)   \\
$g_{7/2}^*$		& { 3.2(2)}&{ 4.9(2) }&{ 3.27(16) }&{ 4.4(2) }&8		& & { 2.73(2)}&{ 2.56(3)}&{ 2.32(2)}&{ 2.20(3)}\\
$h_{11/2}$	& 12.5(6)& 11.1(6)&11.6(5) &12.7(6) &12	& &1.17(2)&1.12(2)&1.14(2)&1.08(2)\\
 \end{tabular}
 \end{ruledtabular}
 \end{table*}

The occupancies of single-proton orbitals are available from  previous measurements of proton removal using the ($d$,$^3$He) reaction. Reference \cite{Wild}, which reports reactions on $N=82$ nuclei from Xe through to Sm, is broadly in agreement with a contemporaneous study on Ba, Ce and Nd \cite{Jones}. A more recent study has been made of Xe and Ba nuclei \cite{Jonathan} with higher  precision. Here we adopt the $^{138}$Ba occupancies from Ref.~\cite{Jonathan} and those for $^{140}$Ce, $^{142}$Nd and $^{144}$Sm from Ref.~\cite{Wild}. 

The pattern of proton occupancies is illustrated in Figure~\ref{occ}, showing significant occupation of the $g_{7/2}$ and $d_{5/2}$ orbitals. The occupancy of the $g_{7/2}$ orbital increases until  $Z=58$, beyond which the changes in occupancy are mainly in the  $d_{5/2}$  orbital. Other orbitals are filled to less than 10\%. The $h_{11/2}$ orbital gradually increases in population across the isotopes, but remains small. Evidence for a low level of occupancy of the $s_{1/2}$ orbital by protons has been found in all nuclei, except for $^{140}$Ce where only an upper limit is available.  The proton occupancy of the $d_{3/2}$  orbital begins to be observable in the two heaviest systems. Although the population of  low-$\ell$ single-proton states are small, they can have a significant effect on the energies of certain neutrons where the orbital overlap is large. 

Calculations of the changes in effective single-neutron energies presented here were performed using the effective two-body force from Reference~\cite{Hosaka} (labelled here as HKT) which was deduced from a G-matrix treatment of the Paris nucleon-nucleon interaction.  The results obtained with that force are very similar to those done using  the phenomenological Schiffer and True \cite{ST} interaction.  Both used single-particle wave functions from infinite oscillator potentials. Individual matrix elements were calculated using the computer code of Reference~\cite{code}, proton-neutron monopole shifts were constructed (these are available as part of the Supplemental Information \cite{supplemental}) and the changes in neutron single-particle energy across the $N=81$ nuclei were obtained using the proton occupancies described above.

\begin{figure}[h!]
\centering
\includegraphics[width=0.5\textwidth]{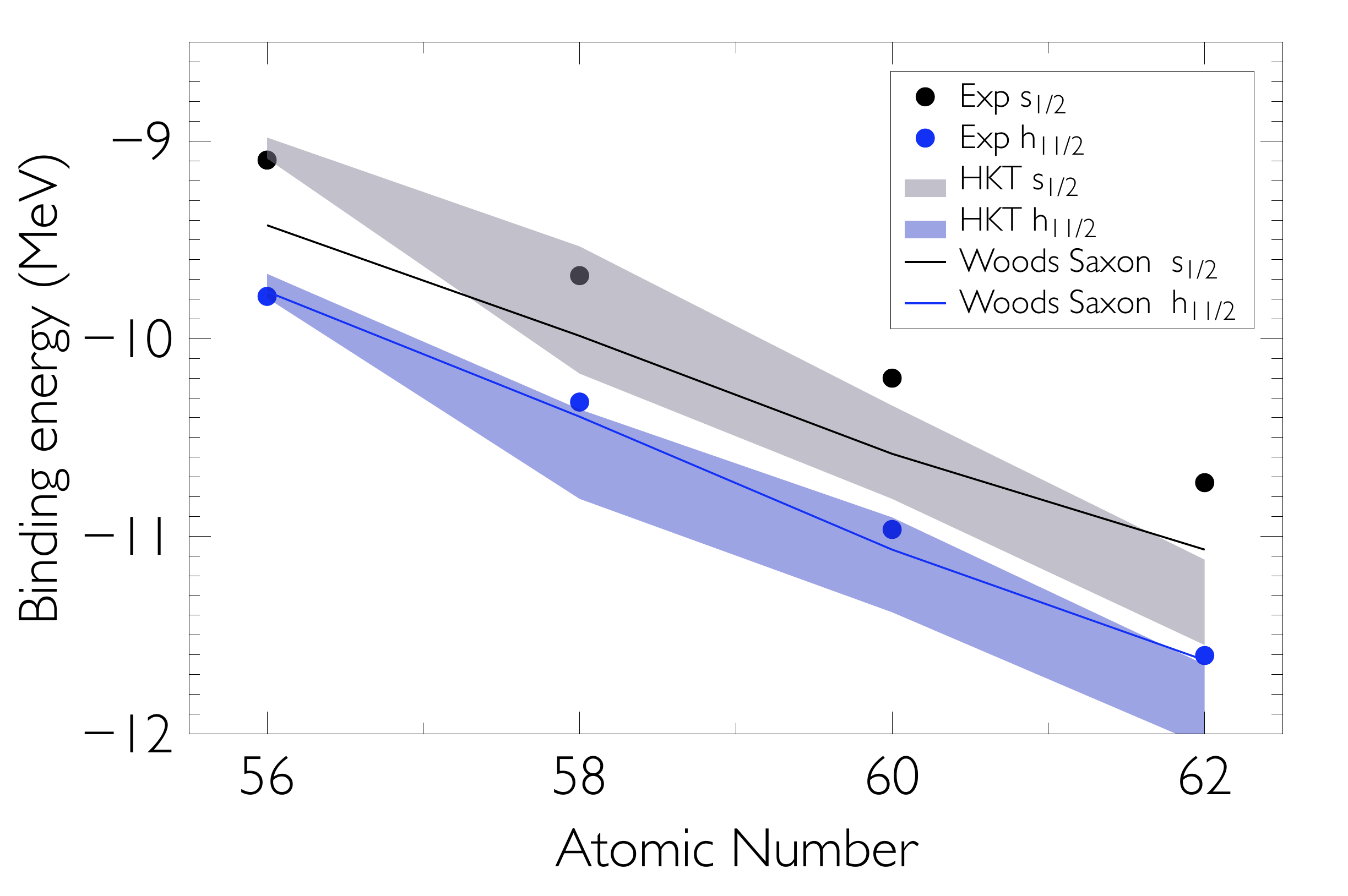}
\caption{\label{BE} Experimental single-particle binding energies for the neutron $s_{1/2}$ (black) and $h_{11/2}$ (blue) orbitals, deduced from the centroids of hole excitation energies.  Calculations used the effective two-body interaction (HKT) of Ref.~\cite{Hosaka} and  proton valence occupancies from Refs.~\cite{Jonathan, Wild}.  These are shown as bands reflecting the uncertainties in the proton occupancies and the absolute value of these calculations along the vertical axis in the figure was shifted to fit the experimental points (see text for more details). The solid lines are Woods-Saxon calculations with standard radius and asymmetry terms with parameters fitted to the $11/2^-$ state in Ba. }
\end{figure}

 To study the effect of the proton occupancy on the relative changes in neutron binding as a function of proton number across the isotopes studied, the experimental data (solid dots) are plotted in Figure~\ref{BE}. A smooth increase in the binding energy of the neutron $s_{1/2}$ and $h_{11/2}$ orbitals is found when adding protons, due to the trends in proton occupancy shown in Figure~\ref{occ}, and the fact that many of the monopole terms have a similar amplitude. Consequently, the effective energy follows that of an averaged global trend of an attractive proton-neutron interaction. Since some of the two-body interactions are different, the change in binding was calculated using the monopole shifts with the HKT interaction and the experimental proton occupancies. Since only the variation with $A$ is meaningful, the absolute value of these calculations along the vertical axis in the figure was shifted to fit the experimental points.  These calculations, including the experimental uncertainties in the proton occupancies, are represented by the shaded areas. (Additionally, the two-body matrix elements themselves are subject to some uncertainty. This is rather difficult to estimate, but is likely of the order of 10\%). 

The monopole shifts for neutron states are particularly sensitive to uncertainties in the occupancy of the corresponding proton orbital due to their large overlap. This is compounded in the case of Ce where only an upper limit on the $s_{1/2}$ proton occupancy had been determined. Indeed, the case of $s_{1/2}$ may be more complicated if some of the weak unassigned strength in the proton-removal reactions is in reality $\ell=0$; for example, there is unassigned strength in the $^{136}$Ba($d$,$^3$He) reaction that amounts to around 0.1 protons (see Table VIII in Ref.~\cite{Jonathan}). 

The trend in the  energy of the neutron $h_{11/2}$ orbital appears reasonably well reproduced by the calculations, as shown in Figure~\ref{BE}, but the slope of the neutron $s_{1/2}$ orbital  is less well predicted in the calculations using monopole shifts from the HKT interaction with harmonic oscillator wave functions. The difference in slope in Figure~\ref{BE} between the data and the monopole-shift calculations for the neutron $s_{1/2}$ orbital suggests that other effects are playing a role for that single-particle state. 

The two-body matrix elements yielding the monopole shifts were calculated using single-particle wave functions in an infinite harmonic oscillator potential where the ordering of the different states is fixed. However, any potential with finite binding is subject to geometric effects such that the single-particle states behave somewhat differently depending on their binding energy relative to the height of the binding potential including the centrifugal term (and Coulomb effects where relevant).  Such effects are known; for instance, they were demonstrated  in Fig 2.30 of Ref.~\cite{B&M} where different neutron orbitals in the 50-82 shell have different behaviors as a function of $A$, notably the $s_{1/2}$ state, and this was discussed in more detail  in Ref.~\cite{Hoffman}. 

The mean field is a sum of two-body interactions, but it is not easy to separate effects that depend on angular momentum  (such as the tensor interaction) from those caused by geometric effects from finite binding. It is therefore instructive to also compare the data to Woods-Saxon calculations, where geometric effects are included, but the angular-momentum dependence from the two-body interaction is not. Fig. 9  shows the results of such calculations with standard radius and asymmetry terms, with parameters fixed to the binding energy of the 11/2$^-$ state in $^{137}$Ba. Such calculations do appear to better reproduce the slope of the $s_{1/2}$ data.

Given these limitations, the level of agreement between data and monopole-shift calculations displayed in Fig.~\ref{BE} is probably reasonable, and constitutes a check on how well the changes in binding energies across the isotopes can be reproduced by the effect of microscopic interactions.

The interpretation of experimental centroids in terms of monopole-shift calculations presented above is a  coarse comparison and it would be useful to understand the fragmentation of single-neutron hole strength across states in the populated nucleus. The general distribution of transfer strength revealed here is reasonably well reproduced by particle-vibration coupling calculations performed a number of years ago \cite{heyde}, given the limitations of the model used (see Fig.~\ref{strength_distribution}). The strong low-lying $\ell=0, 2$ and 5 strength is well reproduced and, although the level of fragmentation is lower than observed due to the restrictions in the model space used, smaller fragments of strength are predicted at higher excitations. The $\ell=4$ strength is predicted to be higher-lying and fragmented, as observed, but any state-to-state correspondence between the experimental data and calculated strength is difficult due to the extent of the fragmentation seen in the experiment.

It would be interesting to compare the strength distributions with the results from modern large-scale shell-model calculations. However, the dimensions of the model space in such a large shell are currently rather difficult to manipulate, making such calculations tricky. Some shell-model calculations have been made around $A=130$ nuclei \cite{teruya}, which includes $^{137}$Ba as one of the heaviest systems considered. Pair-truncated shell-model calculations have been discussed for $^{137}$Ba and $^{139}$Ce \cite{higashiyama}. The results in both cases have so far only been compared to level energies and  electromagnetic moments; predictions of spectroscopic factors are not readily available in the literature. We hope that the current data will inform large-scale calculations as they become available in the future.

In summary, neutron-hole strength in the $N=81$ nuclei $^{137}$Ba, $^{139}$Ce, $^{141}$Nd and $^{143}$Sm has been studied in the ($p$,$d$) and ($^3$He,$\alpha$) neutron-removal reactions at energies of 23 and 34~MeV, respectively. Relative spectroscopic factors extracted through a DWBA analysis and centroids of single-particle strength have been established. The majority of the strength has been observed for the $s_{1/2}$ and $h_{11/2}$ orbitals. Strong fragmentation of strength was observed for the  $g_{7/2}$ orbital, which is more deeply bound and significant strength lies outside of the measured excitation energy range. It proved difficult to properly disentangle $d_{3/2}$ and $d_{5/2}$ strength;  the combined $\ell=2$ strength distribution is broad and also seems to suffer from unobserved, presumably $d_{5/2}$, fragments. Changes in the effect of monopole shifts of neutron energies due to changes in proton occupancy appear to reproduce the trends in the effective single-particle energies of the $s_{1/2}$ and $h_{11/2}$ orbital, at least given the influence of a number of other effects on the former orbital.


\begin{acknowledgments}
We are grateful to John Greene (Argonne National Laboratory) for his careful preparation of the $N=82$ targets used in this work and to the staff at Yale for their assistance in running the experiments. This work was supported by the UK Science and Technology Facilities Council and the US Department of Energy under contract numbers DE-FG02-91ER-40609 and DE-AC02-06CH11357.
\end{acknowledgments}


\end{document}